# Broadband Low-Frequency Near-Perfect Sound Absorber via Coupled Metasurfaces

Keqiang Lyu[1], Mohamed Farhat[2], Ying Wu[1,2]*

[1]Physical Science and Engineering (PSE) Division, King Abdullah University of Science and Technology (KAUST), Thuwal 23955-6900, Saudi Arabia

[2]Computer, Electrical and Mathematical Science and Engineering (CEMSE) Division, King Abdullah University of Science and Technology (KAUST), Thuwal 23955-6900, Saudi Arabia

*Contact author: ying.wu@kaust.edu.sa

We propose a simple yet effective method for low-frequency broadband acoustic absorption. The absorber consists of two concentric space-coiling resonators with distinct resonance frequencies, with the inner resonator characterized by a low-quality factor ($Q$) and the outer resonator by a high $Q$-factor. The coupling between the two resonators enables efficient broadband absorption within a deep-subwavelength regime of over $\lambda/15$. Numerical simulations, theoretical analysis, and experimental measurements demonstrate that highly efficient (>80%) low-frequency broadband absorption is achieved in the range of 198–315 Hz, as well as a 60% fractional bandwidth spanning 183–334 Hz. Furthermore, with the outer dimension fixed, adjusting the parameters of the resonator enables flexible tuning of the absorption band across a broad frequency range. This work presents a powerful design methodology that eliminates the need for traditional complex spatially arranged multi-resonator assemblies. By solely employing a single class of resonant units, thin and efficient broadband absorbers can be achieved, offering various application prospects in the field of low-frequency sound absorption.

The suppression of low-frequency broadband noise is a focal topic in acoustics and engineering due to its applications in building acoustics and environmental noise control. This type of noise is particularly challenging to suppress because of its long wavelength and weak inherent dissipation. Conventional absorbers such as micro-perforated panels (MPPs) [1–4], porous media, and/or fibrous materials [5–7] operate under the causality limit [6,8], which requires large thicknesses to achieve strong absorption especially at the lower frequencies [1,9]. Consequently, such designs result in structures much thicker than the corresponding wavelength, rendering them unsuitable for compact applications. In recent years, the emergence of acoustic metamaterials [10–13] and metasurfaces [14–17] opened new avenues for designing low-frequency absorbers with deeply subwavelength thickness. Various configurations were designed, including membrane-type resonators [16,18–21], Helmholtz resonators [22–25], split-tube resonators [26–28], coherent perfect absorbers [29–31], Fabry–Pérot channels [32,33], and space-coiling structures [26,27,33–42]. However, the absorption bandwidth of these devices remains narrow owing to the inherently resonant nature of the meta-structures.

To tackle the bandwidth limitation, several strategies were put forward for designing broadband low-frequency absorbers, e.g., multilayer systems achieved by stacking Fabry–Pérot channels with distinct absorption peaks [32,43], hybrid structures combining MPPs with cavities of varying depths [44–48], and compact configurations realized by assembling different Helmholtz resonators with perforated plates [49–54]. Compared with these approaches, space-coiling structures were shown to be particularly effective in extending the propagation path within thin structures and enhancing sound absorption at lower frequencies. In fact, the narrowing of the channels intensifies the thermoviscous effects at the air-wall interface, thereby enabling efficient low-frequency absorption with smaller cavity volumes. Inspired by this design strategy, several geometric variants were developed in the form of reflective acoustic metasurface, such as circular [26,38–40], square [34,35], elliptical [27] and gradient design [42]. Even so, a single space-coiling resonator (SCR) remains insufficient for broadband absorption within the constraint of a thin design. A common solution is to stack multiple SCRs [33,42,55,56] or combine them with MPPs [57] and/or porous materials [58]. Nevertheless, such coupled absorbers often involve many components or complex assembly procedures.

To achieve an optimal balance among absorption efficiency, operating bandwidth, structural thickness, and design simplicity, we propose in this Letter a low-frequency broadband reflective metasurface based on SCRs backed by a rigid wall. The design is based on multi-resonant coupling between two concentric resonators with distinct resonant frequencies, one at low and the other at high frequency. This absorber exhibits a fractional bandwidth of 60% in the range of 183–334 Hz, with an average absorption coefficient above 80% between 198 and 315 Hz. The underlying broadband absorption mechanism is investigated in detail, and temporal coupled-mode theory (TCMT) is employed to quantitatively analyze the effects of combining low-$Q$ and high-$Q$ SCRs for constructing an efficient low-frequency broadband sound absorber. The functionality is experimentally validated using impedance-tube measurements based on the two-microphone method and shows good agreement with numerical and theoretical predictions. This work paves the way for practical implementations of ultrathin structures for low-frequency sound absorption.

To mitigate low-frequency airborne noise below 300 Hz (i.e., wavelength above 1 m in air), we propose, as shown in Fig. 1(a), a sound absorber composed of periodic resonant units backed by a rigid wall, where $L$ is the period of the array and $d$ denotes the gap between the units and the wall. The enlarged view in Fig. 1(b) illustrates a cross-sectional view of the space-coiling unit, consisting of a central circular air cavity surrounded by two concentric SCRs with different geometrical features and resonance frequencies. Each SCR comprises several identical spiral segments separated by thin solid walls. The inner resonator (orange region) contains two identical channels, while the outer one (green region) is divided into three parts, forming different folded channels, with outer and inner radii $R_1$ and $r_2$, respectively. The entire structure is constructed from photosensitive epoxy, which is a commonly used 3D printing material. The structural parameters are selected as follows: $L = 99.58$ mm, $d = 10$ mm, $R_1 = 47.43$ mm, $w_1 = 5.5$ mm, $t_1 = 1.43$ mm, $r_2 = 22.75$ mm, $w_2 = 0.99$ mm, $t_2 = 1.98$ mm and $D = 94.86$ mm. The mass density and sound speed of air are $\rho = 1.21$ kg/m$^3$ and $c = 343$ m/s, respectively.

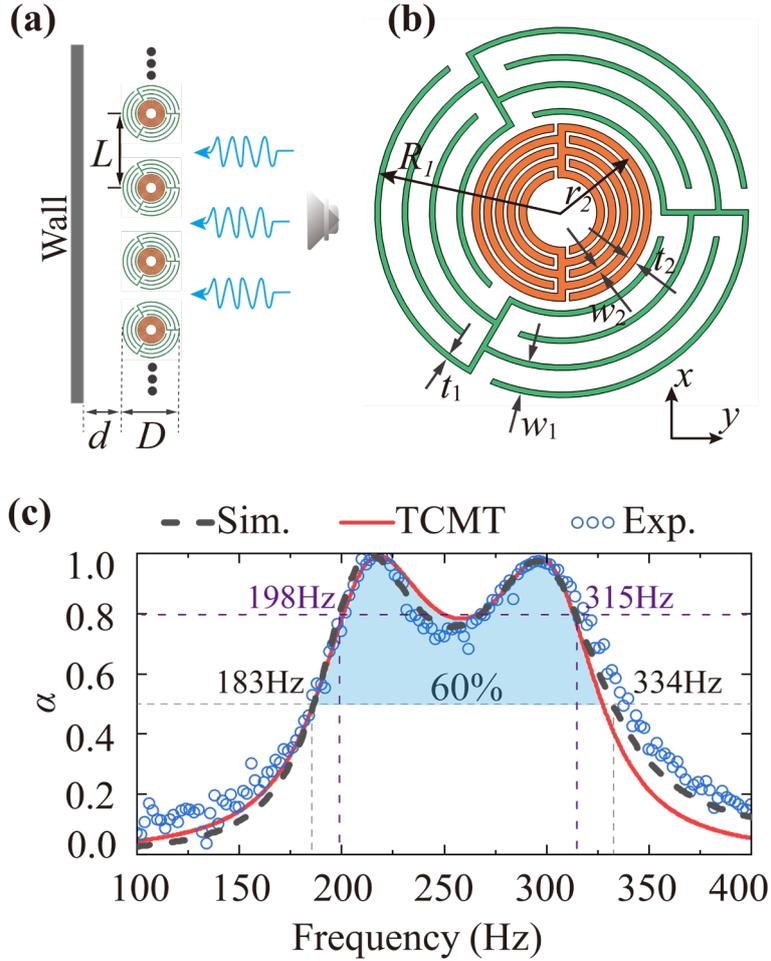

Fig. 1. **Low-frequency SCR absorber.** (a) Schematic of the proposed low-frequency sound absorber with SCR units backed by a rigid wall. (b) The enlarged view reveals the central air cavity and two concentric SCRs forming different folding paths. (c) Sound absorption coefficient $\alpha$ from experiment (blue circles), simulation (black dashed curve), and TCMT (red solid curve), with high absorption ($\alpha > 0.8$) in the range of 198 Hz and 315 Hz and a 60% semi-absorption bandwidth (blue-shaded region).

To numerically investigate the proposed absorber, we employ the frequency-domain solver of COMSOL Multiphysics' thermoviscous acoustic module to account for both thermal and viscous energy dissipation in the narrow channels in parallel with the pressure acoustics module in the surrounding to reduce the computation cost. The walls of the absorber (black lines surrounding the green and the orange solids shown in Fig. 1(b)) are assumed as rigid in the simulations because of the large impedance mismatch between air and the used solid materials. The dynamic viscosity of air is $\mu_d = 1.85 \times 10^{-5}$ Pa·s, its thermal conductivity $k = 0.0258$ W/(m·K), and its heat capacity at constant temperature is $C_p = 1005.4$ J/(kg·s). As shown in Fig. 1(c), the sound absorption coefficient $\alpha$ of the proposed SCR-based metamaterial is evaluated by comparing experimental measurement (blue circles), finite-element simulations (black dashed curve), and TCMT (red solid curve), all of which exhibit good agreement. The proposed metamaterial achieves near-perfect sound absorption ($\alpha > 98\%$) at 213 Hz and 296 Hz, with a semi-absorption bandwidth (for which $\alpha \geq 0.5$) reaching 60% (highlighted by the blue shaded region in Fig. 1(c)). Furthermore, at the frequency corresponding to the first absorption peak (213 Hz), the wavelength of the incident sound is more than 15 times the metamaterial thickness ($d + D$), demonstrating near-perfect absorption at a deep subwavelength regime.

To understand the underlying design principle and mechanism, we employ TCMT, which provides a powerful framework for describing the absorption characteristics of single or multiple resonators directly from their intrinsic resonance parameters, without reference to a specific geometry. In an open system such as ours (shown in Fig. 1(a)), a resonant mode such as that of the SCR is characterized by the leakage quality-factor $Q_{\text{leak}}^{-1}$ and the loss quality-factor $Q_{\text{loss}}^{-1}$, corresponding to the ratio of the stored energy to the leaked and dissipated energy, respectively. Combining these contributions, the overall quality-factor ($Q$) of the SCR can be readily expressed as $Q^{-1} = Q_{\text{leak}}^{-1} + Q_{\text{loss}}^{-1}$. Moreover, $Q_n^{-1} = \Delta\omega_n/\omega_n$, where $\Delta\omega_n$ denotes the half-maximum absorption bandwidth, $\omega_n$ is the angular resonant frequency, and $n = a, b$ denote the outer and inner resonator, respectively. According to TCMT, the sound absorption coefficient $\alpha$ under normal incidence in a one-port system is given [33,59–61] by:

$$\alpha_n = \frac{4Q_{\text{leak},n}^{-1} Q_{\text{loss},n}^{-1}}{4(\omega/\omega_n - 1)^2 + (Q_{\text{leak},n}^{-1} + Q_{\text{loss},n}^{-1})^2}, \quad (1)$$

In our design concept, the outer and inner SCRs are engineered to possess relatively lower and higher loss factors, respectively. Specifically, the outer SCR (shown in green in Fig. 2(a)) is designed with a wider channel width, yielding a simulated resonant frequency of $f_a = 213$ Hz, $Q_{\text{leak},a}^{-1} = 0.3$, and $Q_{\text{loss},a}^{-1} = 0.038$ (see the Supplemental Material [62] for detailed derivation and calculation of these parameters). $Q_{\text{loss}}^{-1}$ primarily originates from the frictional dissipation within the folded channel, whereas $Q_{\text{leak}}^{-1}$ is associated with the inevitable leakage of sound energy from the folded channel into the surrounding medium. As shown in Fig. 2(a), the absorption results obtained from TCMT (red solid curve) are in good agreement with the finite-elements method (FEM) simulations (black circles), especially in the vicinity of the absorption peak, validating the accuracy of the theoretical model. In Fig. 2(b), the narrower channel width $w_1$ leads to an inner SCR characterized by $Q_{\text{leak},b}^{-1} < Q_{\text{loss},b}^{-1}$ (see the Supplemental Material [62]). The simulated resonant frequency is $f_b = 248$ Hz, with $Q_{\text{leak},b}^{-1} = 0.028$ and $Q_{\text{loss},b}^{-1} = 0.2037$. Likewise, the absorption spectrum predicted by TCMT (red solid curve) closely reproduces the FEM (black circles) results. Furthermore, the variations of the leakage and loss factors of both the inner and outer SCRs with respect to different geometric parameters are also investigated in the Supplemental Material [62].

When the inner and outer SCRs are brought together, their mutual coupling provides an effective mechanism for broadband absorption. To illustrate this effect, Fig. 2(c) shows the dependence of the absorption coefficient on frequency and the inner SCR radius $r_2$. As $r_2$ increases, the resonance frequency of the inner SCR undergoes a redshift, leading to stronger coupling with the outer SCR and ultimately yielding an optimal bandwidth, as indicated by the red dashed double-arrow. As shown in the Supplemental Material [62], we also investigate the absorption performance of the absorber under different rotation angles of the inner and outer SCRs.

It is found that rotating the inner SCR has a negligible influence on the absorption bandwidth, while rotating the outer SCR within 0–30° and 80–120° maintains a perfect absorption coefficient above 0.8 in the Supplemental Material [62]. Furthermore, we analyzed the effects of geometric parameters such as the unit spacing $L$ and the distance $d$ between the absorber and the rigid backing wall. The variations in $L$ and $d$ were found to affect the absorption bandwidth, which can be attributed to the change in coupling strength between adjacent SCRs.

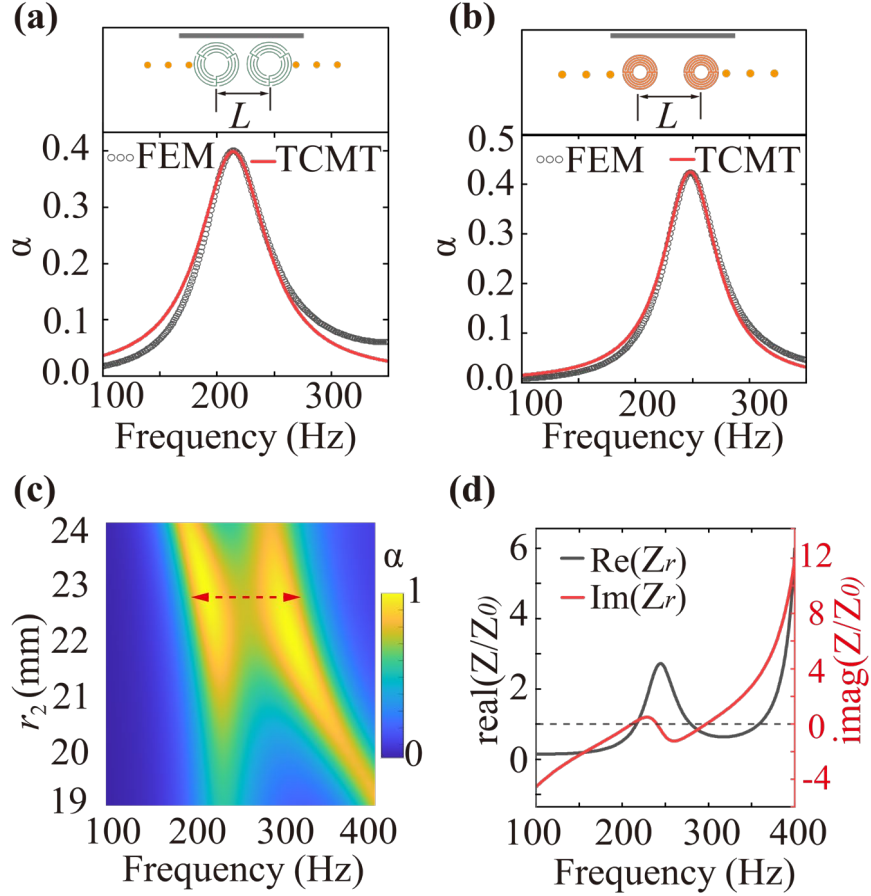

Fig. 2. **Absorption performance of the combined-SCRs metasurface.** Absorption of the (a) outer SCR and (b) inner SCR showing good agreement between FEM simulations and TCMT analysis. The insets show the schematics of the SCRs. (c) Coupling effect between the two SCRs, where tuning the inner radius enables optimization of the absorption bandwidth. (d) Surface impedance matching between the metasurface and air, with the imaginary part (red curve) approaching zero and the real part (black curve) near-unity.

From the impedance perspective, Fig. 2(d) illustrates the mechanism underlying the high absorption. The relative surface impedance $Z_r = Z/Z_0$ is plotted as a function of frequency. The surface impedance can be calculated using $Z = p_s/v_s$, where $p_s$ and $v_s$ denote the surface sound pressure and the normal particle velocity in the simulation, respectively, and $Z_0 = \rho_0 c_0$ is the specific acoustic impedance of air. As seen in Fig. 2(d), the imaginary part of the normalized impedance crosses zero at exactly 213 Hz and 296 Hz, consistent with the absorption peaks observed in Fig. 1(b). Meanwhile, the corresponding real parts remain close to unity (0.99 and 0.98), indicating that the absorber is near-perfectly impedance-matched to the surrounding air at resonance. Such near-ideal impedance matching suppresses reflection and thus enhances energy coupling into the SCRs, thereby enabling highly efficient sound absorption (the transmission channel is completely suppressed owing to the rigid backing wall).

In order to gain a quantitative understanding of the coupling scenarios, we formulate the TCMT model for the coupled SCRs configuration (see the Supplemental Material [62]), where each SCR is described by the governing equations, given as follows [33,59–61]:

$$\frac{d\psi_1}{dt} = (-i\omega_1 - \gamma_1 - \Gamma_1)\psi_1 - i\kappa_c\psi_2 + \sqrt{2\gamma_1}S_+, \quad (2)$$

$$\frac{d\psi_2}{dt} = (-i\omega_2 - \gamma_2 - \Gamma_2)\psi_2 - i\kappa_c\psi_1, \quad (3)$$

$$S_- = -S_+ + \sqrt{2\gamma_1}\psi_1, \quad (4)$$

where, $\omega_{1,2}$ are the eigenfrequencies of the outer and inner SCR corresponding to $f_1 = 253\text{Hz}$ and $f_2 = 260\text{Hz}$. Here, $\psi_{1,2}$ denotes the complex amplitude of each resonator, while $S_+$ and $S_-$ represent the incident and reflected waves, respectively. $\gamma_{1,2} = Q_{\text{leak},1,2}^{-1}\omega_{1,2}/2$ and $\Gamma_{1,2} = Q_{\text{loss},1,2}^{-1}\omega_{1,2}/2$, with $\gamma_{1,2}$ and $\Gamma_{1,2}$ representing the radiative and dissipative decay rates, and $Q_{\text{leak},1,2}^{-1} = Q_{\text{leak},a,b}^{-1}$, $Q_{\text{loss},1,2}^{-1} = Q_{\text{loss},a,b}^{-1}$, and $\kappa_c$ is their mutual coupling factor. As detailed in the Supplemental Material [62], the total reflectance $R = \left|\frac{S_-}{S_+}\right|^2$ can then be expressed as

$$R = \left|1 - \frac{2\gamma_1}{-i(\omega - \omega_1) + \gamma_1 + \Gamma_1 + \frac{\kappa_c^2}{-i(\omega - \omega_2) + \gamma_2 + \Gamma_2}}\right|^2. \quad (5)$$

and the absorption coefficient is $\alpha = 1 - R$. The good agreement between the theoretical absorption curve (red curve) and the simulated result (black dashed curve) in Fig.1(c) confirms that the TCMT accurately captures the coupled behavior of the two SCRs.

To gain deeper insight into the origin of the broadband absorption mechanism, it is noted that a single SCR cannot achieve high absorption over a wide frequency range. Broadband absorption requires the absorber to exhibit both strong field enhancement and significant thermoviscous losses at multiple frequencies. Figure 3(a) shows the thermoviscous power dissipation as a function of frequency, with separate contributions from the total, outer, and inner SCR. Two prominent peaks are observed, which correspond to the strong absorption modes of the metasurface. The dissipation curves indicate that the inner SCR (blue curve) contributes significantly more to the total energy loss than the outer SCR (red curve).

The top panels of Fig. 3(b) show that at the first absorption peak frequency, i.e., 213 Hz (marked by the red star in Fig. 3(a)), the acoustic field of the absorber is predominantly confined within the inner SCR, and both the inner and outer SCRs vibrate in phase. As further discussed in the Supplemental Material [62], at this frequency the resonance of the outer SCR dominates, corresponding to a monopolar mode in which the acoustic energy is localized mainly within the central cavity of the outer SCR, while the inner SCR exhibits a weak response. In this case, neither resonator alone can provide strong absorption.

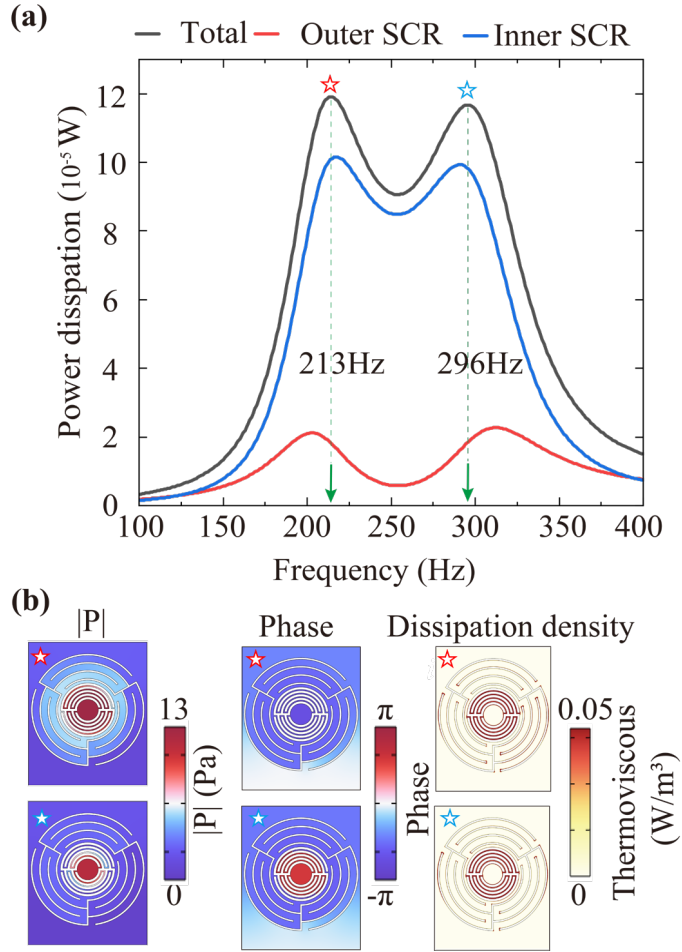

**Fig. 3. Acoustic field characteristics of the combined-SCRs metasurface.** (a) Thermoviscous power dissipation as a function of frequency, with separate contributions from the total (black curve), outer (red curve), and inner channels (blue curve). (b) Magnitude |P| (left panels), phase (middle panels) and dissipation density (right panels) of the SCRs at the two absorption peak frequencies, i.e., 213 Hz (marked by the red star) and 296 Hz (marked by the blue star).

However, when the two SCRs are combined, the strongly confined field generated by the outer SCR within its central cavity is captured by the inner SCR and efficiently dissipated through thermoviscous losses in its narrow channels. This process indicates that the field enhancement is mainly provided by the outer SCR, whereas the energy dissipation responsible for absorption originates from the inner SCR, at this frequency.

The bottom panels of Fig. 3(b) show that at the second absorption peak frequency, i.e., 296 Hz (marked by the blue star in Fig. 3(a)), the acoustic energy of the absorber also remains concentrated within the channels of the inner SCR, where the inner and outer SCRs oscillate out of phase. As discussed in the Supplemental Material [62], the resonance of the outer SCR is weak at this frequency and insufficient to provide strong absorption, whereas the inner resonator governs the response in a monopolar mode, with the acoustic field localized inside its channels. Unlike the first absorption peak, when the two SCRs are combined at this frequency, both the field enhancement and thermoviscous losses mainly originate from the inner SCR, at this frequency.

In addition, the right panels of Fig. 3(b) illustrate the spatial distributions of thermoviscous power density, revealing that the energy dissipation is mainly concentrated within the narrow channels, thereby validating the absorption mechanism.

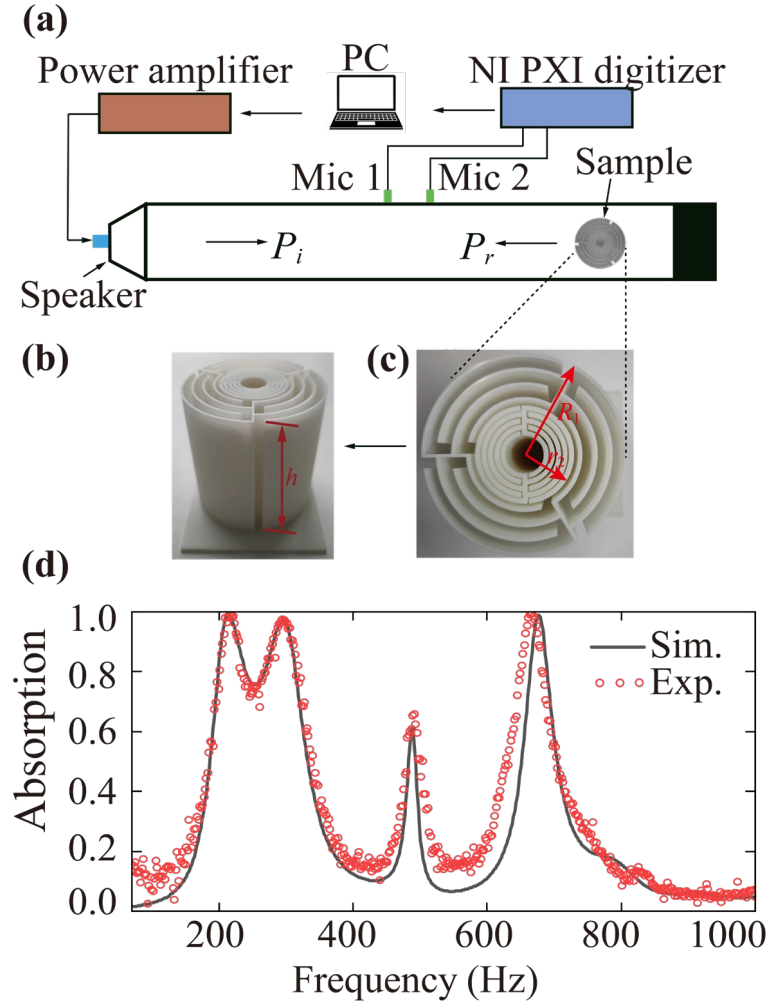

Fig. 4. **Experimental validation of the proposed metasurface.** (a) Schematic of the setup for absorption measurements using an impedance tube. (b) Photograph of the fabricated sample with overall height $h = 10$ cm. (c) Top view of the sample showing the coiled channel with outer and inner SCR radii, $R_1$ and $r_2$. (d) Comparison of simulated and measured absorption spectra for a broadband frequency range (100-1000 Hz).

For the experimental validation of the low-frequency broadband absorption performance, we measured the spectrum in an acoustic impedance tube, as illustrated in Fig. 4(a). The setup comprises a loudspeaker mounted at the tube entrance, two microphones to capture the pressure field [63], and auxiliary electronics including a signal generator and power amplifier. The sample was placed at the closed end of the impedance tube and backed by a rigid block. Figure 4(b) depicts the photograph of the fabricated absorber, produced by 3D printing using photosensitive epoxy (density 1180 kg/m³, longitudinal wave speed 2720 m/s, and shear wave speed 1460 m/s), with a total height of $h = 10$ cm.

Detailed parameter settings are provided in the Supplementary Material [62]. The corresponding cross-sectional view in Fig. 4(c) shows the outer and inner resonator radii ($R_1$ and $r_2$). During the measurement, the computer-generated acoustic signal was delivered through a data acquisition card and power amplifier to the loudspeaker, producing the incident sound wave. The pressure field was recorded by two microphones positioned at fixed locations inside the tube (as can be seen in Fig. 4(a)). Based on the acoustic pressure measured at these two points, the incident and reflected waves were determined, from which the absorption coefficient of the sample was subsequently obtained.

As shown in Fig. 4(d), the experimental results (red circles) exhibit good agreement with the simulated absorption spectrum (black solid line) across the frequency range of 100–1000 Hz. In particular, the close match between 213 Hz and 296 Hz confirms the accuracy of the present design and the numerical simulation. Some discrepancies appear in the non-resonant frequency regions, which may arise from imperfections introduced during the 3D printing and material preparation processes. Surface roughness, in particular, tends to increase the absorption coefficient outside the resonance band, leading to the small deviations observed in the figure. Overall, the proposed absorber demonstrates a pronounced advantage in low-frequency absorption.

In addition, we investigated the tunability and robustness of the proposed absorber. Under fixed external dimensions, broadband perfect absorption can be freely tuned across different frequency ranges (160–435Hz) by designing units with different coiled channel configurations, without the need of parallel multiple units with varying geometrical parameters [62]. Moreover, the absorber maintains stable performance under oblique incidence (within $50°$), exhibiting absorption exceeding 50% even at large incident angles, which highlights its robustness against angular variations, as detailed in the Supplemental Material [62]. Finally, we compared the sound absorption characteristics of representative low-frequency reflection-type metasurfaces based on space-coiling configurations reported over the past decade (see the Supplemental Material [62]). The comparison focuses on the normalized thickness ($D/\lambda$) and fractional bandwidth ($\alpha>50\%$) of metasurfaces with single-type structural configurations. The results reveal that the proposed metasurface achieves the widest fractional bandwidth (60%) in the deep subwavelength region, surpassing conventional broadband absorbers that typically rely on coupling multiple resonators operating at different frequency ranges.

In conclusion, we have demonstrated a broadband low-frequency acoustic absorber based on coupled concentric space-coiling resonators, achieving two near-perfect absorption peaks at 213 Hz and 296 Hz with a deep-subwavelength thickness of $\lambda/15.4$, a fractional bandwidth of 60%, and an average absorptance above 80%. The absorption originates from the interplay of thermoviscous dissipation in the narrow channel of the inner SCR, radiation leakage of the outer SCR and their hybrid coupling, as validated by TCMT, simulations, and experiment measurements. Further analyses confirm the robustness of the absorber against geometric variations, tunability of broadband absorption frequency under fixed external dimensions, and stable absorption performance within 50° incidence angles. These results highlight the potential of our design for compact, efficient, and practical low-frequency noise control.

## ACKNOWLEDGMENTS

KL acknowledges fruitful discussions related to this work with Rfaqat Ali, Lijuan Fan, He Liu, Fathi, Iman, Long Sun. The work described here is supported by the Office of the Sponsored Research (OSR) at King Abdullah University of Science and Technology (KAUST) under grant No. ORFS-CRG11-2022-5055, ORFS-OFP-2023-5560, and BAS/01/1626-01-01.

## DATA AVAILABILITY

The data that support the findings of this article are not publicly available. The data are available from the authors upon reasonable request.